\documentclass[]{article}
\usepackage{latexsym}
\begin{document}
\titlepage
\begin{flushright}
T03/133\\
hep-th/0309117 \\
\end{flushright}
\vskip 1cm
\begin{center}
{\large \bf Detuned Branes and Supersymmetry Breaking}
\end{center}

\vspace*{5mm} \noindent 

\centerline{ Ph. Brax\footnote{brax@spht.saclay.cea.fr}${}^{a}$, N. Chatillon\footnote{chatillon@spht.saclay.cea.fr}${}^{a}$}

\vskip 0.5cm 
\centerline{\em ${}^{a}$ Service de Physique Th\'eorique} 
\centerline{\em CEA/DSM/SPhT, Unit\'e de recherche associ\'ee au CNRS,}
\centerline{\em CEA-Saclay F-91191 Gif/Yvette cedex, France} 
\vskip 2cm

\begin{center}
{\bf Abstract}
\end{center}

We consider the spontaneous breaking of supersymmetry in
five dimensional supergravity with boundaries where the
supersymmetry  breaking mechanism is provided by the even part of
the bulk prepotential. The supersymmetric action comprises
boundary brane terms with detuned tensions. The two branes have
opposite tensions. We analyse the possible vacua with
spontaneously broken supersymmetry. A class of solutions
corresponds to rotated branes in an $AdS_5$ bulk. In particular
parallel  branes which are rotated with respect to the bulk
preserve $1/4$ of supersymmetry. We  analyse more general vacua
using the  low energy effective action for gravity coupled to the
radion field. Supersymmetry breaking implies that the radion field
acquires a potential which is negative and unbounded from below.
This potential is modified when coupling the boundary branes to a
bulk four--form field. For a brane charge larger than the deficit
of brane tension,  the radion potential is bounded from below
while remaining flat when the charge equals the deficit of brane
tension.

%

%

\section{Introduction}
Supersymmetry breaking is one of the stumbling blocks of physics
beyond the standard model. String inspired models suffer from the
impossibility of breaking supersymmetry at the perturbative level
\cite{Polchinski_bk}. Non-perturbative mechanisms such as gaugino
condensation provide a useful framework \cite{gaugino}, although
one has to resort to race-track potentials which are not free from
fine-tuning problems \cite{racetrack}.  An appropriate way of
breaking supersymmetry would have to face stringent experimental
constraints springing from both particle physics and cosmology. In
cosmology, the cosmological constant issue requires to adjust the
vacuum energy to almost zero energy compared to the Planck scale
\cite{SNIa}. This is difficult to reconcile with supergravity
breaking where one expects  a vacuum energy of the order
$O(m_{3/2}^2M_{\rm Pl }^2)$. In particle physics, one expects that
the non-observation of superpartners is due to a large mass
splitting of the order of one TeV within supersymmetry multiplets.

Recently a new paradigm for supersymmetry breaking has been
provided by extra--dimensional models \cite{susybreak_extradim}.
In particular D-branes \cite{D-branes} allow new mechanisms of
supersymmetry breaking. For instance supersymmetry breaking may be
transmitted from one brane to another by bulk fields leading to
hierarchical energy scales \cite{bulk_transmission}. One can also
design brane anti-brane configurations which break all
supersymmetry \cite{sen,brane_antibrane, Brax:2001xf} and
non-supersymmetric branes which tend to be unstable objects
\cite{sen,unstable_branes}. A $N=1$ superspace formulation of 5d supergravity in a flat background with boundary branes
has been recently presented \cite{phil}.

In this paper we will focus on a simpler framework, i.e. the
supersymmetric Randall-Sundrum model where bulk supergravity is
coupled to two boundary branes
\cite{RS1,RS2,Altendorfer:2000rr,Falkowski:2000er,kallosh,offshell,flp2,flp3,susy_radion,
Brax:2001xf,
susyRS_arbitrary_tensions,Brax:2002vs,Lalak:2002kx,twisted_sugra,twist_warp_sugra}.
In particular we will be interested in spontaneously broken
supersymmetry whereby branes become tilted, i.e branes which are
not hyperplanes at constant $x_5$ when the bulk metric reads
\begin{equation}
ds^2= dx_5^2 + a^2 (x_5) \eta_{\mu\nu} dx^\mu dx^\nu.
\end{equation}
In section 2 we will consider the supersymmetric Randall-Sundrum
model spontaneously  broken  by the even part of the bulk
prepotential \cite{Brax:2001xf}. This amounts to detuning the
tensions of the boundary branes by a factor $T<1$. In section 3 we
present a  class of solutions representing rotated branes. These
branes are hyperplanes at angles.  In the non-singular case of
rotated and parallel brane configurations,  Lorentz invariance is
broken down to $SO(1,2)$.  In section 4 we find the Killing
spinors for  parallel branes preserving $1/4$ of supersymmetry. We
also consider the massive  gravitini showing that they do not
survive due to the incompatibility between the boundary conditions
on the branes and the propagation in the bulk. Finally in section
5  we analyse supersymmetry breaking from the four dimensional
point of view using the low energy effective action.  This allows
one to discuss generic supersymmetry breaking vacua.  We find that
the radion potential is negative and unbounded from below. Hence
cosmological solutions all end up in a big crunch singularity. We
also discuss non-singular static configurations.  Finally we
couple the boundary branes to a bulk four--form field. One finds a
BPS bound linking the charge to the deficit of brane tension. When
the charge exceeds the BPS bound the radion potential becomes
bounded from below with a vacuum corresponding to infinitely far
apart branes. The BPS case when the charge equals the deficit of
brane tension leads to a vanishing potential for the radion field.
In section 6 we give some comments and conclusions. We have also
included an appendix where the equivalence between a single
$AdS_4$-- brane and a rotated brane has been made explicit.
Another appendix deals with details about massless gravitons and
rotated branes.

\section{Supergravity with Tilted Boundary Branes}
\subsection{Boundary conditions on tilted branes}
We  extend the known supersymmetrization of the Randall-Sundrum
model \cite{Falkowski:2000er} to the case where the boundary
branes are tilted. In particular we  deduce a supersymmetric
Lagrangian and associated boundary conditions which are compatible
with arbitrarily  curved boundary branes. The issue of
supersymmetry breaking by the tilting of the boundary branes will
be tackled in section 3.

Let us consider pure supergravity in the bulk
comprising a supergravity multiplet containing the metric
$g_{ab}$, $a,\ b =1\dots 5$, the gravitini $\psi_a^A$ where
$A=1,2$ is a symplectic index (the gravitini are
Majorana-symplectic spinors), and finally the graviphoton $A_a$.
The graviphoton can be used as the gauge field when gauging a
$U(1)_R$ subgroup of the $SU(2)_R$ symmetry. Such a gauging implies
the presence of a cosmological constant $-6k^2$ in the bulk and
modifies the supersymmetry variations (with the normalization of
\cite{ovrut})
\begin{equation}
\delta \psi_a^A\supset  D_a \epsilon^A  -\frac{i\sqrt 2}{3}{\cal
P}^{AB}\gamma_a\epsilon_B .
\end{equation}
The bulk bosonic  action reduces to the $AdS_5$ action
\begin{equation}
S=\frac{1}{\kappa_5^2}\int d^5x \sqrt{-g}( \frac{1}{2}R+6k^2)
\end{equation}
admitting a bulk solution of the equations of motion in  the form
\cite{RS1}
\begin{equation}
ds^2=dx_5^2 +a^2(x_5) \eta_{\mu\nu}dx^\mu dx^\nu
\end{equation}
where $a(x_5)=e^{-kx_5}$ is the warp factor.

We would like to define supergravity on a $Z_2$ orbifold with
arbitrary boundary branes. To simplify matters we assume that the
boundary branes are located at
\begin{equation}
x_5^1=\xi_1(x^\mu),\ x_5^2= \rho + \xi_2(x^\mu)
\end{equation}
where $\xi_{1,2}$ are infinitesimal displacements.

In order to
define a covering space it is convenient to perform a change of
coordinates which straightens up both branes defining
\begin{equation}
\hat x^5=x^5 + \xi ^5 (x^5,x^\mu)
\end{equation}
where the function $\xi^5$ must satisfy
\begin{equation}
\xi^5 ( \xi_1,x^\mu)=-\xi_1,\ \xi^5 (\rho +\xi_2,x^\mu)= -\xi_2 .
\end{equation}
The resulting branes are located at
\begin{equation}
\hat x_1^5=0,\ \hat x^5_2=\rho .
\end{equation}
The simplest possibility for $\xi^5$  is provided by a linear function in $x^5$.
One must accompany the change in $x^5$ with a small change of four
dimensional coordinates \cite{straightened_branes}
\begin{equation}
\hat x^\mu= x^\mu - \eta^{\mu\nu}\Big(\int_0^{x^5}
\frac{du}{a^2(u)}\partial_\nu \xi^5(u,x^{\lambda})\Big)\equiv
x^\mu +\xi^\mu .
\end{equation}
Such a transformation implies that the metric is not in a normal
Gaussian form anymore
\begin{equation}
\hat g_{55}= 1- 2\partial_5 \xi_5
\end{equation}
where $\xi_5=\xi^5$ and
\begin{equation}
\hat
g_{\mu\nu}=a^2\Big[(1+2k\xi^5)\eta_{\mu\nu}+2\Big(\int\frac{du}{a^2(u)}\partial_\mu\partial_\nu
\xi^5(u,x^{\lambda})\Big)\Big] .
\end{equation}
The main property of the change of coordinates where the branes
are straightened up is that
\begin{equation}
\hat g_{\mu 5}=0 .
\end{equation}
Defining the radion as the distance between the boundary branes
\begin{equation}
R=\int_0^\rho \sqrt{\hat g_{55}}dx_5 = \rho+\xi_2-\xi_1
\end{equation}
one finds that it is independent of the choice of $\xi^5$.

We can now define the $Z_2$ action in the $(\hat x^5,\hat x^\mu)$
coordinates by compactifying the $\hat x^5$ coordinate on a circle
of perimeter  $2\rho$ and identifying $(\hat x^5,\hat x^\mu) \equiv (-\hat x^5,\hat x^\mu)$.
The two boundary branes are now fixed points under the $Z_2$
action.

In the non-gaussian coordinate system, one can impose  the usual boundary
conditions on the
fields which acquire the usual parity under the $Z_2$ action. Thus
$\hat g_{\mu\nu}$ and $\hat g_{55}$ are even, $\hat g_{\mu 5}$ is
odd, $\hat A_\mu$ is odd while $\hat A_5$ is even. The gravitini
have particular boundary conditions
\begin{equation}
(\delta^A_B -\Gamma{\cal Q}^A_{\ B})\hat \psi_\mu^B=0
\end{equation}
and
\begin{equation}
(\delta^A_B +\Gamma {\cal Q}^A_{\ B})\hat \psi_5^B=0
\end{equation}
on both branes where we have introduced
\begin{equation}
\Gamma=-n^a\gamma_a
\end{equation}
and $n^a$ is the normalized normal vector to the brane whose fifth
component is chosen to be negative. We have also used a matrix
${\cal Q}^A_{\ B}$ in the Lie algebra of $SU(2)$. Finally, the odd
fields $\hat g_{\mu 5}$ and $\hat A_\mu$ vanish on the branes.

We can now perform the inverse change of coordinates going back to
the situation where the branes are tilted and the bulk admits
Gaussian normal coordinates. As already stated the $g_{\mu 5}$
component of the metric vanishes in both frames. This  leads to no
restriction on the metric from the $Z_2$ symmetry. We will focus
on the boundary conditions for the gravitini
in the supergauge $A_5=0$ and $\psi_5^A=0$.
Indeed we will analyse the mass spectrum of Kaluza-Klein states
for the gravitini using this constraint in
section 4. The boundary condition of the graviphoton on the brane
becomes
\begin{equation}
A_\mu\vert_{brane} =0
\end{equation}
while the two boundary conditions for the gravitini become
\begin{equation}
(\delta^A_B -\Gamma{\cal Q}^A_{\ B})\psi_\mu^B\vert_{brane}=0
\end{equation}
supplemented with
\begin{equation}
(\partial_5 \xi^\nu ) \psi_\nu^A\vert_{brane}=0 \label{bound2} .
\end{equation}
Here the normal vector reads
\begin{equation}
n^a_{1,2}=\frac{1}{\sqrt{\tilde n^{1,2}_a g^{ab}\tilde n^{1,2}_b}}\tilde n^a_{1,2}
\end{equation}
where
\begin{equation}
\tilde n_a^{1,2}= (\partial_\mu \xi_{1,2},-1) .
\end{equation}
These boundary conditions will be crucial in determining the spectrum of gravitini.
Finally the boundary condition for the spinors parametrizing the supersymmetric variations
is
\begin{equation}
(\delta^A_B -\Gamma {\cal Q}^A_{\ B})\epsilon^B=0 \label{bound}
\end{equation}
on the brane.
\subsection{Supersymmetric action}
Let us come back to the supersymmetrization of the brane and bulk
system. To do that we need to specify the prepotential in terms of
even and odd functions under the $Z_2$ symmetry \cite{Brax:2001xf}
\begin{equation}
{\cal P}^{AB}=i(\frac{9}{8})^{1/2}[\epsilon (\hat x_5)kT {\cal
Q}^{AB}+ k\sqrt{1-T^2} { \Sigma}^{AB}]
\end{equation}
where the matrices ${\cal Q}$ and $\Sigma$ anticommute $\{{{\cal
Q},\Sigma}\}=0$ and ${\cal Q}^2=1,\ \Sigma^2=1$, together with
the Lie algebra properties of hermiticity and tracelessness. The
function $\epsilon (\hat x_5)$ jumps from -1 to 1 at the origin
and from 1 to -1 at $\rho$.

The variation of the action under a
supersymmetry transformation of spinor $\epsilon^A$ leads to
boundary terms from the gravitino kinetic terms. More precisely
the action of the fifth derivative on the odd part of the
prepotential leads to delta--functions which can be compensated by
the variation of the boundary action \cite{Brax:2001xf,
susyRS_arbitrary_tensions}
\begin{equation}
S=-\frac{6kT}{\kappa_5^2}\int_1 d^4x
\sqrt{-h}+\frac{6kT}{\kappa_5^2}\int_2 d^4x \sqrt{-h}
\end{equation}
where $h_{\mu\nu}$ is the induced metric on the brane. The brane
tensions are constant and opposite
\begin{equation}
\lambda_{\pm}=\pm \frac{6kT}{\kappa_5^2}
\end{equation}
Notice that the factor $T<1$ introduced in the prepotential and
parameterizing its even part leads to a detuning of the brane
tension with respect to the tuned Randall-Sundrum case $T=1$. More
general detunings in supergravity have been considered in
\cite{susyRS_arbitrary_tensions}.

The action comprising the bulk supergravity Lagrangian and the
boundary brane terms is supersymmetric. We will see that as soon
as $T<1$ supersymmetry is spontaneously broken. The boundary brane
action with $T<1$ implies that static branes sitting at constant
$x_5$ in the bulk are no longer vacua of the theory.

\section{Rotated Branes and Lorentz Invariance Violation}
\subsection{Rotated branes}
One can easily find a class of  vacua of the spontaneously broken
models $T<1$.
Going to conformal coordinates
\begin{equation}
ds^2= a^2(y)(dy^2 + \eta_{\mu\nu}dx^\mu dx^\nu)
\end{equation}
in the bulk where $a(y)=1/ky$, one can accomodate the change of
boundary conditions induced by the detuning of the brane tensions
by rotating the boundary branes with respect to the $y$ axis.
Decompose four dimensional vectors according to an orthogonal
basis $(h^\mu, e^\mu_i)$ where $h^\mu$ is space--like and $e^\mu_i
e^\nu_j\eta_{\mu\nu}=\eta_{ij}$. For instance\footnote{ We denote
by dot the Minkowski scalar product ; similarly $h^2$ is defined
as $h_\mu \eta^{\mu\nu} h_{\nu}$.} we identify $x^\mu=
X\frac{h^\mu}{\sqrt {h^2}} +X^i e_i^\mu$ and  $dx^\mu dx_\mu=
-dX_0^2 + dX_1^2 +dX_2^2 +dX^2$. Consider the rotation
\begin{eqnarray}
\hat y&=& \frac{y-\sqrt{h^2} X}{\sqrt{1+h^2}}\nonumber \\
\hat X &=& \frac{X+\sqrt {h^2} y}{\sqrt{1+h^2}} \nonumber \\
\end{eqnarray}
where the angle of rotation $\theta$ satisfies $\tan \theta=
\sqrt{h^2}$. We leave $(X_0,X_1,X_2)$ invariant. This leads to the
metric
\begin{equation}
ds^2=a^2(\frac{\hat y+\sqrt{h^2} \hat
X}{\sqrt{1+h^2}})(d\hat y^2 -dX_0^2 + dX_1^2 +dX_2^2 +d\hat X^2)
\end{equation}
in the bulk. One of the boundary branes located at $(\hat
y=constant)$ satisfies the Israel boundary conditions
\begin{equation}
\frac{\partial_{\hat y} a}{a^2}\vert_{\hat
y=constant}=-\frac{k}{\sqrt{1+h^2}}= -kT
\end{equation}
provided the tilting vector $h^\mu$ fulfills
\begin{equation}
h^2 =\frac{1}{T^2}-1 .
\end{equation}
Notice that this vector is indeed space-like. Perform then the
inverse rotation. The brane is located at $y= h.x +constant$ in
conformal coordinates $(x^\mu,y)$. Apply the same procedure to the
second boundary brane. We thus find that the two boundary branes
depend on two tilting vectors $h_{1,2}^\mu$ with the same norm. We
choose the positions of the  branes to be at
\begin{equation}
y=h_1.x+\frac{1}{k},\ y=h_2.x + \frac{e^{k\rho}}{k} .
\label{cuckoo}
\end{equation}
The constants have been chosen for convenience.  The branes are
hyperplanes in conformal coordinates. They correspond to rotated
branes which are moving if $h_0\ne 0$. In the following we will
refer to this brane configuration as rotated branes. We will be
particularly interested in the parallel configurations
\begin{equation}
y=h.x+\frac{1}{k},\ y=h.x + \frac{e^{k\rho}}{k} .
\end{equation}
where $h=h_1=h_2$. Such configurations are not perturbatively
connected to the usual boundary branes of the Randall--Sundrum
model located at $y=const.$ Indeed the distance between the
branes, i.e. the radion, is
\begin{equation}
R(x)=\frac{1}{k}\ln\big ( \frac{h.x +
\frac{e^{k\rho}}{k}}{h.x+\frac{1}{k}}\big )
\end{equation}
which goes to infinity on the planes where the two branes reach
$y=0$.

 As $T<1$ one might have expected  to find the
$AdS_4$ brane configurations of \cite{dewolfe}. This is only
possible if the two boundary branes  carry two tensions of the
same sign. Here the opposite sign of the tensions forces us to
consider two rotated branes with the same norm $h^2_1=h^2_2$
\cite{Brax:2001xf} (see appendix A for more details).

Let us examine the geometry of one of the boundary branes in more
detail. Notice first that the boundary branes hit the boundary
$y=0$ along an codimension two plane. Focus on the positive
tension brane. This happens for $h_1.x+ \frac{1}{k}=0$. For
convenience sake, let us focus on the case where $h_1$ lies along
the third axis only. The induced metric reads
\begin{equation}
ds_4^2= dz^2 + e^{-2\sqrt{1-T^2}kz}(-dt^2 +dx_1^2 +dx_2^2)
\label{indu}
\end{equation}
where we have identified
\begin{equation}
k\sqrt{1-T^2}z= \ln (kh_1 x_3+1)
\end{equation}
Notice that such a coordinate system is only valid on one of the
branes at a time and not globally.  When the brane hits the
boundary at $y=0$, the coordinate $z\to -\infty$. Moreover $z$
covers the whole real line. Using the coordinates $(t,x_1,x_2,z)$
the boundary geometry does not present any pathological behaviour
and is simply identified with $AdS_4$ with cosmological constant
\begin{equation}
\Lambda_{AdS_4}= -3k^2 (1-T^2)
\end{equation}
The same analysis applies to the second brane of negative tension.

There are three typical global configurations that one may
envisage for the two brane system. First of all the branes can
intersect and the supergravity description breaks down at the
intersection hyperplane. This is the generic case. One may also
consider the case where the would-be intersection is beyond the
$AdS_5$ boundary located at $y=0$.  Finally one may send  the
would-be intersection to infinity by choosing the two branes to be
parallel.  In the following we will focus on the parallel case.

\subsection{Lorentz invariance violation}

One can use either the description in terms of rotated  branes  or change coordinates to non-rotated  boundary branes with
a space-time dependent metric.
 In order to study the violation of Lorentz invariance
it is convenient to use the
non-rotated boundary brane picture after the change of coordinates
\begin{equation}
\hat x^5= \frac{\rho}{\rho+\xi_2-\xi_1}(x^5 -\xi_1) .
\end{equation}
For small brane tilting $\xi_{1,2}$ this corresponds to
\begin{equation}
\xi^5=\frac{(\xi_1-\xi_2)}{\rho}x_5 -\xi_1
\end{equation}
where (\ref{cuckoo}) leads to
\begin{equation}
\xi_1= h.x,\ \xi_2= e^{-k\rho}h.x .
\end{equation}
for small $\xi_{1,2}$ and small enough $x$. The metric is not in the normal Gaussian
form anymore
\begin{equation}
\hat g_{55}= 1+ 2\frac{\xi_2-\xi_1}{\rho}
\end{equation}
and
\begin{equation}
\hat g_{\mu\nu}= a^2 (1+2k\xi_5)\eta_{\mu\nu} .
\end{equation}
This implies that the background can be described as
obtained from the Randall-Sundrum background by switching on
a non-trivial vev for the radion field
\begin{equation}
<R>(x)=\rho +\xi_2(x) -\xi_1(x) .
\end{equation}
This is an explicitly Poincar\'e violating effect as it is
$x^\mu$-dependent.
More precisely, let us perform a Lorentz transformation
\begin{equation}
\tilde x^\mu= x^\mu +M^{\mu\nu}x_\nu
\end{equation}
where $M^{\mu\nu}$ is antisymmetric.
The bulk metric is invariant provided
\begin{equation}
M^{\mu}_{\ \nu} h^\nu=0 .
\end{equation}
The Lorentz group is broken to $SO(1,2)$.
Similarly translation invariance $\tilde x^\mu=x^\mu +a^\mu$ implies
that $a_\mu h^{\mu}=0$, i.e. momenta $p^\mu$ are only defined in the plane orthogonal to
$h$
\begin{equation}
p_\mu h^{\mu}=0 .
\end{equation}
The Poincar\'e invariance of the bulk metric is  the three dimensional
Poincar\'e group.

Another way of realizing that Poincar\'e invariance is broken
springs from the induced metric (\ref{indu}) on one of the boundary branes. The
boundary metric of $AdS_4$ admits a $SO(2,3)$ invariance with a
$SO(1,2)$ subgroup  of linearly acting Lorentz transformations.

\subsection{Brane supersymmetry breaking}

As well as Lorentz invariance violation in the bulk, supersymmetry on the branes is also
affected.
We consider brane matter in the form of four
dimensional chiral superfields $\Phi$ living on the positive
tension brane. The coupling of brane matter to the radion
superfield springs from  the K\"{a}hler
potential \cite{flp3}
\begin{equation}
K=-3\ln (1-e^{-k(S+\bar S)} -\vert \Phi\vert^2)
\end{equation}
where $\Phi$ is dimensionless and $S +\bar S =2R$. Now the kinetic terms of the radion
field in the Einstein frame reads
\begin{equation}
L_{kin}=K_{S\bar S}\vert\frac{\partial S}{\partial x}\vert^2 .
\end{equation}
This leads to a soft supersymmetric breaking mass
for the chiral multiplets
\begin{equation}
m^2_{\Phi}= 6k^2e^{-k(S+\bar S)}\vert\frac{\partial S}{\partial
x}\vert^2
\end{equation}
for large $R$.
More explicitly, the soft breaking mass reads
\begin{equation}
m^2_{\Phi}= 6k^2 \frac{1-T^2}{T^2}e^{-2k<R>} .
\end{equation}
As expected this mass term is Poincar\'e  violating due to the explicit
dependence of $<R>$ on the space-time coordinates.

\subsection{Massless gravitons}

We will analyse  the influence of the breaking of Poincar\'e
invariance on massless gravitons.  Let us first state the Israel
junction condition on both branes
\begin{equation}
K_{\mu\nu}-h_{\mu\nu}K=  \pm 3Tkh_{\mu\nu}
\end{equation}
where $h_{\mu\nu}$ is the induced metric on the branes.
This implies that
\begin{equation}
K_{\mu\nu}=\mp Tkh_{\mu\nu} .
\end{equation}
It is convenient to introduce the vielbein $e^a_\mu=
\frac{\partial Y^a}{\partial x^\mu}$ where $Y^a$ is the embedding
of the branes in conformal coordinates. In our case the only
non-zero components are
\begin{equation}
e^a_\mu=\delta^a_\mu,\ e^5_\mu= \partial_\mu Y^5\equiv h_\mu .
\end{equation}
The extrinsic curvature tensor is defined by
\begin{equation}
K_{\mu\nu}=e^a_\mu e^b_\nu D_a n_b .
\end{equation}
The perturbed metric is chosen in the bulk
\begin{equation}
ds^2_5= a^2(y)\Big(dy^2
+(\eta_{\mu\nu}+n_{\mu\nu}(x^{\lambda},y))dx^\mu dx^\nu\Big) .
\end{equation}
This implies that to first order in $n_{\mu\nu}$
\begin{equation}
n=aA(h_\mu,-1)
\end{equation}
where
\begin{equation}
A=\frac{1}{\sqrt{1+h_\mu \eta^{\mu\nu}h_\nu}}(1+
\frac{1}{2}\frac{h_\mu
n^{\mu\nu}h_\nu}{1+h_\nu\eta^{\mu\nu}h_\nu}) .
\end{equation}
The perturbed Israel junction conditions to first order in
$n_{\mu\nu}$ read
\begin{equation}
\partial_5 n_{\mu\nu}+h_\mu\partial_5n^\rho_\nu h_\rho +
h_\nu\partial_5n^\rho_\mu h_\rho+
h_\lambda(\partial_\mu n_\nu^\lambda +\partial_\nu n^\lambda_\mu
-\partial^\lambda n_{\mu\nu})=ak\frac{h_\mu n^{\mu\nu}h_\nu}{1+h_\mu\eta^{\mu\nu}h_\nu}(
\eta_{\mu\nu}+h_\mu h_\nu)
\label{israel}
\end{equation}
where we have explicitly used $\partial_\mu h^\nu=0$.

A sufficient way of satisfying the  junction conditions  is to impose  Neumann
boundary conditions
\begin{equation}
n^a\partial_a n_{\mu\nu}=0
\label{i1}
\end{equation}
on the branes.
In the bulk the
graviton modes read
\begin{equation}
n_{\mu\nu}(y,x^{\mu})=\phi(y) e^{ip.x}n_{\mu\nu}
\end{equation}
where $p^2=0$ for massless excitations  and $n_{\mu\nu}$
is transverse and traceless.
In the massless case $\phi(y)$ is constant so that the Neumann
boundary condition is automatically satisfied provided that
\begin{equation}
p.h=0
\end{equation}
corresponding to the absence of translation invariance in the normal
direction
to the brane.
The boundary condition implies then that
\begin{equation}
h^\mu n_{\mu\nu}=0 .
\label{cond1}
\end{equation}
This is tantamount to saying that gravitons must be polarised
along the brane, i.e. orthogonal to the brane normal vector.
The necessary and sufficient analysis of the boundary conditions is detailed in Appendix B.

The condition $\partial_\mu h_\nu =0$
breaks the residual gauge symmetry $x^\mu \to x^\mu +\epsilon^\mu$
where $\Box \epsilon^\mu =0$ to preserve the transversality of the
metric perturbation  and
 $\partial_c\epsilon^c=0$ to preserve its tracelessness.
We have to impose
\begin{equation}
\epsilon.h=0
\end{equation}
reducing the number of  gauge degrees of freedom.

The tensor $n_{\mu\nu}$ has five independent components.
Let us choose
$h=(0,0,0,h^3)$. By an appropriate $SO(1,2)$ Lorentz transformation one can put
$p=(\omega,0,\omega,0)$. The transversality condition imposes that $n_{0\mu}=n_{2\mu}$.
Moreover (\ref{cond1}) leads to $n_{3\mu }=0$. The residual gauge invariance
with $\epsilon^3=0$ allows one to put $n_{00}=n_{01}=0$.
The polarisation matrix vanishes altogether.
One can also understand this result by noticing that the polarisation
condition and the restricted gauge invariance imply that massless
gravitons
effectively behave like $(1+2)$ massless gravitons. In three
dimensions,
a traceless transverse symmetric matrix has two independent degrees of
freedom
which are eaten by the two residual gauge invariances.

In conclusion we have seen that rotated and parallel  branes are classical
vacua
of five dimensional supergravity with a non vanishing even part of the
prepotential.
Poincar\'e invariance is broken down to three dimensional Poincar\'e
invariance
implying that no massless gravitons propagate in the bulk.

\section{Supersymmetry Breaking}
In this section we will examine the spontaneous breaking of supersymmetry
due to the detuning of the brane tension and the corresponding rotation
of the boundary branes with respect to the $AdS_5$ bulk.
We will first consider the Killing spinors and then the gravitini
in the bulk.
\subsection{Killing Spinors}
The brane configuration preserves  supersymmetry provided one can
find Killing spinors compatible with the boundary conditions due
to the $Z_2$ orbifold symmetry \cite{kallosh}. Imposing that the
supersymmetric variation $\delta\psi_a^A=0$, $a=1\dots 5$, $A=1,2$
implies that
\begin{equation}
D_a\epsilon^A-\frac{i\sqrt 2}{3}\gamma_a{\cal P}^A_{\
B}\epsilon^B=0
\end{equation}
where we are considering the bulk metric in conformal coordinates.
This reduces to
\begin{equation}
\partial_a\epsilon^A + \frac{ka}{2} \gamma^5_{\ a} \epsilon^A +\frac{ka}{2}\gamma_a(
T{\cal Q}^A_{\ B} + \sqrt{1-T^2}\Sigma_{\ B}^A
)\epsilon^B=0\label{kill}
\end{equation}
written with flat  space Dirac matrices now. Let us introduce
the projectors
\begin{equation}
\Omega_{\pm B}^{A}=\frac{1}{2}[\delta^A_B\pm\gamma_5(T{\cal
Q}^A_{\ B} + \sqrt{1-T^2}\Sigma_{\ B}^A )]
\end{equation}
and the eigenspinors
\begin{equation}
\Omega_{\pm B}^A\epsilon_\pm^B=\pm \epsilon_{\pm}^A .
\end{equation}
Now projecting (\ref{kill}) for $a=5$ on both chiralities and
considering the following ansatz
\begin{equation}
\epsilon^A_\pm (x^\mu , y)=f_\pm (ky)\epsilon^A_\pm (x^\mu)
\end{equation}
yields
\begin{equation}
\frac{df_\pm }{dy}\pm\frac{1}{2y}f_{\pm}=0
\end{equation}
and therefore the profile of the Killing spinors
\begin{equation}
f_{\pm}(y)=a^{\pm 1/ 2}(y) .
\end{equation}
Notice that the two types of Killing spinors are either peaked on
the first or on the second brane.
Let us now use the remaining Killing spinor equations for $a=\mu$
\begin{equation}
\partial_{\mu}\epsilon^A-ka(y)\gamma_\mu\gamma^5\Omega_{-B}^A\epsilon^B=0
..
\end{equation}
Projecting again on both chiralities and using the explicit
dependence of $\epsilon_\pm$ on $y$, we find that
\begin{eqnarray}
\epsilon_-(x^\mu)&=&\epsilon^0_-
\nonumber \\
\epsilon_+(x^\mu)&=&\epsilon^0_+ + k x^{\mu} \gamma_{\mu}\gamma^5
\epsilon^0_-
\end{eqnarray}
where $\epsilon^0_{\pm}$ are space-time independent and we have
omitted the symplectic indices. The corresponding four dimensional
Majorana spinors are defined as follows :
\begin{equation}
\epsilon_+^1=f_+(y)\epsilon^+_R(x^\mu)\ ,\ \epsilon_+^2=f_+(y)
\epsilon^+_L(x^\mu)
\end{equation}
and
\begin{equation}
\epsilon_-^1=f_-(y)\epsilon^-_L(x^\mu)\ ,\ \epsilon_-^2=-f_-(y)
\epsilon^-_R(x^\mu) .
\end{equation}
Let us now examine the compatibility with the boundary condition
(\ref{bound}), evaluated on  both branes and depending on the
matrices $\Gamma=T(\gamma^5-h_\mu\gamma^\mu)$. After a
$SU(2)_R$ rotation sending ${\cal Q}$ in the real
$(\sigma_1,\sigma_3)$ plane and $T{\cal Q}+\sqrt{1-T^2}{\Sigma}\to
\sigma_3$, we find that ${\cal Q}^1_{\ 1}=T$ and ${\cal Q}^1_{\
2}=\varepsilon \sqrt{1-T^2},\ \varepsilon=\pm$.
 Defining the normalized vector $u_\mu=
\frac{T}{\sqrt{1-T^2}}h_\mu$ if $T\ne 1$, we find that the boundary condition
reduces to
\begin{equation}
f_-(y)\epsilon^-(x^\mu)=-\varepsilon
\frac{\sqrt{1-T^2}}{T}\frac{1+\varepsilon
u.\gamma}{2}f_+(y)\epsilon^+(x^\mu) \label{spacelike boundary
condition}
\end{equation}
evaluated on both branes. This implies
that
\begin{equation}
\epsilon^- (x^\mu)=0 \ , \ (1+\varepsilon u.\gamma
)\epsilon^+(x^{\mu})=0 ,
\end{equation}
i.e. only two components of the spinor $\epsilon^+$ survive the
projection.  Thus only $1/4$ of the original eight supersymmetries
are conserved by the vacuum for $T<1$ parallel branes. 
The existence of configurations with less than four supersymmetries in 4d has
already been considered in \cite{gibbons}.
The
explicit appearance of the non-zero $u_{\mu}$ signals a breaking
of 4d Lorentz invariance too.

In the case $T=1$ with non-zero light-like tilting vector
$h_{\mu}$, we have on each brane instead of (\ref{spacelike
boundary condition}) :
\begin{equation}
f_-(y) \epsilon_-(x^{\mu}) =
-\frac{1}{2}(h.\gamma)f_+(y)\epsilon_+(x^{\mu}) .
\end{equation}
This results in the (4d Lorentz violating) constraints
\begin{equation}
\epsilon^- (x^\mu)=0 \ , \ (h.\gamma )\epsilon^+(x^{\mu})=0 .
\end{equation}
This also preserves $1/4$ of the supersymmetries.

\subsection{Gravitini in the bulk}

Having realized that a generic tilted configuration breaks
supersymmetry one can expect that the tower of Kaluza-Klein modes
for the gravitini is going to be affected too. Let us consider the
gravitino equation in the bulk first
\begin{equation}
\gamma^{abc}D_b\psi_c^A+\sqrt 2 i {\cal P}^A_{\ B} \psi_{cB}=0
\end{equation}
where we consider the supersymmetric gauge $\psi_5=0$.
This can be reexpressed as
\begin{equation}
\gamma^{\mu\nu\rho}\partial_\nu\psi_\rho^A
-a\gamma^5\gamma^{\mu\nu}\partial_5\psi_\rho^A
+\frac{5k}{2}a\gamma^5\gamma^{\nu\rho}\psi_\rho^A
-3ka\gamma^5\gamma^{\mu\nu}(\Omega_{+B}^{A} \psi_{\rho}^B)=0
\end{equation}
for $a=\mu$ and the gamma matrices are associated to the
Minkowski metric here.
Let us perform as before a $SU(2)_R$ rotation to send $\Omega_\pm$ to
\begin{equation}
\Omega^A_{\pm B} \to \frac{1}{2}(\delta_B^A\pm \gamma^5(\sigma_3)^A_{\
B})
\end{equation}
and decompose the symplectic spinors according to
\begin{equation}
(\Omega_+ \psi_\rho)^1=\phi_+(x_5)\psi^+_{\rho R}(x^\mu),\
(\Omega_+ \psi_\rho)^2=\phi_+ (x_5) \psi^+_{\rho L}(x^\mu)
\end{equation}
and
\begin{equation}
(\Omega_- \psi_\rho)^1=\phi_-(x_5)\psi^-_{\rho L}(x^\mu),\
(\Omega_- \psi_\rho)^2=-\phi_- (x_5) \psi^-_{\rho R}(x^\mu)
\end{equation}
where the two spinors $\psi^{\pm}$ are four dimensional Majorana spinors.
Due to the symplectic features of the five dimensional spinors,
it is sufficient to consider the $A=1$ equation only.
Moreover we are looking for massive gravitini from the four
dimensional point of view
\begin{equation}
\gamma^{\mu\nu\rho}\partial_\nu\psi^\pm_\rho=m\gamma^{\mu\rho}\psi^{\mp}_\rho
..
\end{equation}
Notice the change of chirality.
The gravitino profiles satisfy
\begin{eqnarray}
\frac{m}{a}\phi_-&=& \partial_5\phi_+ +\frac{k}{2}\phi_+\nonumber \\
\frac{m}{a}\phi_+&=& -\partial_5 \phi_-
+\frac{5k}{2}\phi_- . \nonumber \\
\end{eqnarray}
These are the same differential equations as for the
supersymmetric Randall-Sundrum model \cite{flp2}. Using the $a=5$
equation we obtain the irreducibility constraint
\begin{equation}
\gamma^\mu\psi_\mu^-=0
\end{equation}
for the massless gravitino $\psi^-_{\mu}$.

Let us now consider the compatibility with the boundary condition
\begin{equation}
(\delta^A_B-\Gamma {\cal Q}^A_{\ B})\psi_\mu^B|_{branes} =0 .
\end{equation}
The discussion is exactly the same as for the Killing spinors. For
proportional tilting vectors, $1/4$ of them would  survive the
projection :
\begin{eqnarray}
&\psi_\mu^-=0& \textrm{ and}
\nonumber \\
&(1+\varepsilon u \cdot \gamma )\psi_\mu^+=0& \textrm{ for } T<1
\nonumber \\
&(h \cdot \gamma)\psi^+_{\mu}=0& \textrm{ for } T=1,\  h_\mu \neq
0 .
\end{eqnarray}
Combined with the Rarita-Schwinger equation connecting the two
spinors $\psi_\mu^\pm$ this leads to
\begin{equation}
\psi_\mu^\pm=0
\end{equation}
in the massive case. Only massless gravitini survive in agreement
with the $1/4$ of remaining supersymmetries.

In conclusion we have found that parallel branes are vacua of
spontaneously broken 5d supergravity preserving only $1/4$ of
supersymmetry and 3d Poincar\'e invariance.

\section{The Low Energy Action}

\subsection{Radion instability}
So far we have only considered parallel and rotated brane vacua of
5d supergravity and shown that supersymmetry is broken in that
case. We have also discarded other vacua which lead to brane
intersections where the supergravity description breaks down. In
this section  we will use the low energy effective action
describing the dynamics of the moduli fields, i.e. the graviton
and the radion in the bosonic sector. This allows one to analyse a
wide range of possible vacua including time dependent
configurations. Nevertheless, the rotated branes may be out of
reach of the low energy approximation due to their
non-perturbative nature compared to the Randall-- Sundrum straight
branes. We will analyse the static solutions of the low energy
equations of motion and find configurations which are locally
similar to the rotated branes.

As an example, one can embed the two boundary branes in an
$AdS_5$-Schwarzschild background with black-hole mass $\mu$ and 3d
sectional curvature $q$ to find that the branes move according to
\begin{equation}
H^2= (T^2-1)k^2 -\frac{q}{r^2}+
\frac{\mu}{r^4}
\end{equation}
where $H=\dot r /r$. These backgrounds break supersymmetry too
\cite{kinetic_susy_breaking}. More generally the brane geometry is
described within the projective approach \cite{projective} where
the brane Einstein equations depend on the matter content of each
brane and the projected Weyl tensor. The latter is bulk dependent
and reduces to the black hole mass in the time-dependent setting.
In general, one does not know how to describe the general
solutions of the boundary brane Einstein equations. On the
contrary when the matter content of the branes is a small
perturbation compared to the brane tension of the Randall-Sundrum
$T=1$ case, one can use the moduli space approximation to
understand the dynamics fully.

This description is known to coincide with the projective approach
as long as the deviation of the brane tension  from the BPS
($T=1$) case   is small, i.e. when $\vert T-1\vert << 1$
\cite{proj2}. The fact that $T\ne 1$ leads to a non vanishing
potential at low energy for the radion field. We will analyse the
solutions of the equations of motion and show that either the
configuration ends up in a cosmological big crunch singularity or
the radion vanishes in a locus where the supergravity description
breaks down.

In the moduli space approach, one considers the bulk metric
\begin{equation}
ds^2=a^2(x_5) g_{\mu\nu}(x^\mu) dx^\mu dx^\nu +dx_5^2
\end{equation}
and the branes are located at $0$ and $R(x^\mu)$. Integrating over the
fifth dimension,
the dynamics of the
branes can be described by an effective action involving the
radion and determined by the underlying supergravity K\"{a}hler
potential \cite{flp3,susy_radion}
\begin{equation}
K=-3\ln (1-e^{-k(S+\bar S)})
\end{equation}
where $S$ is the radion superfield and $S+\bar S=2R$ for its
bosonic part.
This specifies the action entirely up to the supersymmetry breaking
potential induced by the detuning of the brane tensions. The potential term is
\begin{equation}
\frac{6k(1-T)}{\kappa_5^2}\int d^4 x \sqrt{-g}(1-e^{-4kR})
\end{equation}
in the brane frame where the Einstein-Hilbert term reads
\begin{equation}
\frac{1}{4k\kappa_5^2}\int d^4x \sqrt{-g} (1-e^{-2kR}){\cal R}
\end{equation}
and ${\cal R}$ is the 4d curvature.
In the Einstein frame the radion action reads
\begin{equation}
-\frac{k}{4\kappa_5^2}\int d^4 x \sqrt {-g}(\frac{6
  e^{2kR}}{k^4 (e^{2kR}-1)^2}(\partial R)^2
  +24(T-1)\frac{(1-e^{-4kR})}{(1-e^{-2kR})^2}) .
\end{equation}
It is convenient to normalise the radion kinetic terms by defining
\begin{equation}
\phi=\frac{1}{2}\ln(\frac{e^{kR}-1}{e^{kR}+1})
\end{equation}
in such a way that the action becomes
\begin{equation}
-\frac{k}{4\kappa_5^2}\int\sqrt{-g}(\frac{6}{k^2}(\partial\phi)^2+12(T-1)\cosh
 (2\phi)) .
\end{equation}
Notice that $\phi$ varies between $-\infty$ and zero.
The potential is unbounded from below and always negative.

Let us take the limit $\phi=0$ where the branes are infinitely far apart, i.e. the one brane system.
The potential reduces to a cosmological constant
\begin{equation}
\Lambda_{\phi =0}=\frac{3(T-1)k}{\kappa_5^2}
\end{equation}
This cosmological constant for a one--brane system coincides with the boundary cosmological constant as obtained from the $AdS_4$ boundary metric
\begin{equation}
\Lambda_{AdS_4}=\frac{3(T^2-1)k^2}{\kappa_4^2}
\end{equation}
for $\vert T-1 \vert << 1$ and we have used $\kappa_4^2=2 k \kappa_5^2$.
The two--brane system has a non trivial potential whose limiting value
is equal to the one--brane cosmological constant.

Let us now show that cosmological solutions of this low energy action
always reach a big crunch singularity. We focus on flat spatial sections.
Looking for FRW solutions in the Einstein frame with scale factor $a$
leads to
\begin{equation}
\frac{\ddot a}{a}=-4 (\frac{(\dot \phi)^2}{k^2}+(1-T)\cosh
(2\phi))
\end{equation}
in 4d Planck units and $\dot {} \  =d/dt$.
Notice that $\ddot a <0$ implying that $\dot a$ decreases. Starting
from $\dot a >0$, the scale factor first increases before reaching a
turning point where $\dot a =0$, then entering a contracting phase
with $\dot a <0$. Eventually the scale factor has to vanish as no
minimum for $a$ can exist. Hence cosmological solutions with a flat
spatial section end up in a big crunch singularity.

Consider static configurations now. We choose the ansatz for the
metric
\begin{equation}
g_{\mu\nu}dx^\mu dx^\nu = dx^2_3 +a^2(x_3) (-dt^2 +d\Omega_{\tilde
k}^2) \label{met}
\end{equation}
where $d\Omega_{\tilde k}^2$ is the line element  of the two dimensional symmetric space with curvature $\tilde k=0,\pm1$.
We also restrict ourselves to $\phi(x_3)$. Define
\begin{equation}
\tilde V= 12k^2 (1-T) \cosh \frac{\tilde \phi}{\sqrt 3}
\end{equation}
where $\tilde \phi= 2\sqrt 3 \phi$ is the canonically normalised
field. Notice that $\tilde V$ is the opposite of the potential appearing
in the action. The Einstein equations lead to the Friedmann
equation
\begin{equation}
H^2= \frac{1}{6} (\frac{\tilde \phi'^2}{2} +\tilde V)
+\frac{\tilde k}{3a^2}
\end{equation}
We have introduced $H= \frac{a'}{a}$ where $'=d/dx_3$. Notice the
factor of 3 in the curvature term as we only consider two
dimensional static symmetric spaces. The Klein-Gordon reads
\begin{equation}
\tilde \phi'' +3H \tilde \phi'= -\frac{\partial \tilde
V}{\partial\tilde \phi}
\end{equation}
We have thus rewritten the equations of motion of static
configurations in a form similar to cosmological equations where
time is replaced by $x_3$.

We are looking for solutions where the branes at  infinity
$x_3=\pm \infty$ in the coordinate system used to define the
metric (\ref{met}) are far away $\phi=0$. This corresponds to the
field starting at the minimum of the potential, the field being
negative decreases and reaches a minimum before falling down to
the origin again. We will consider the case where the minimum is
close to the origin and $\phi'$ remains small. In that case the
Friedmann equation reduces to
\begin{equation}
H^2= -\frac{\Lambda_{AdS_4}}{3} + \frac{\tilde k}{3a^2}
\end{equation}
Choosing $\tilde k=-1$, we find a bounce
\begin{equation}
a(x_3)= \frac{1}{\sqrt{-\Lambda_{AdS_4}}}\cosh
(\sqrt{\frac{-\Lambda_{AdS_4}}{3}}x_3)
\end{equation}
We can now discuss the solutions of the Klein-Gordon equation in
two regimes. When $\vert \sqrt{\frac{-\Lambda_{AdS_4}}{3}}x_3\vert
>> 1$, one can approximate $H^2=\frac{-\Lambda_{AdS_4}}{3}$
and the Klein-Gordon equation becomes
\begin{equation}
\tilde \phi'' +3H \tilde \phi'+ 2H^2 \tilde \phi=0
\end{equation}
Hence  the field $\tilde \phi$ behaves like
\begin{equation}
\tilde \phi = A_{-} e^{\sqrt{\frac{-\Lambda_{AdS_4}}{3}}
x_3} +B_{-} e^{2\sqrt{\frac{-\Lambda_{AdS_4}}{3}}
x_3}
\end{equation}
for $x_3<0$ and
\begin{equation} \tilde \phi = A_{+}
e^{-\sqrt{\frac{-\Lambda_{AdS_4}}{3}}x_3} +B_{+}
e^{-2\sqrt{\frac{-\Lambda_{AdS_4}}{3}}x_3}
\end{equation}
for $x_3 >0$ and some constants $A_\pm<0$ and $B_{\pm}<0$. When
$\vert \sqrt{\frac{-\Lambda_{AdS_4}}{3}}x_3\vert <<  1$, one can
neglect $H$ in the Klein-Gordon equation leading to
\begin{equation}
\tilde \phi = C \cos (\sqrt{\frac{-2\Lambda_{AdS_4}}{3}}x_3)
\end{equation}
for some constant $C$ . It is interesting to connect $\tilde \phi$
and the radion. For large $\vert x_3\vert $ the radion is linear
\begin{equation}
R(x_3) = h \vert x_3 \vert
\end{equation}
while it is goes to a non-vanishing constant for small $x_3$.
Hence we find that the second brane is  wedge--like. Locally this
solution resembles the rotated branes obtained previously although in a different coordinate system.

For phenomenological models, this type of behaviour for the radion
is particularly undesirable as it leads to large Lorentz violating
effects due to the spatial dependence of the radion field. It also
leads to a negative potential energy  as compared to the
observation of a small positive cosmological constant. Of course
this prompts towards modifying the previous setting by including
other sources of supersymmetry breaking which may compensate the
negative radion  potential that we have derived.

\subsection{Charged Branes}

So far we have considered neutral branes. Here we will consider charged branes.
One can also introduce
a bulk four--form $C$ coupling to the branes
\begin{equation}
S=-\frac{1}{2}\int F\wedge *F -\frac{2\kappa_5}{\sqrt 3}Q\int_+ C +\frac{2\kappa_5}{\sqrt 3}Q\int_- C
\end{equation}
where $Q$ is the charge of the positive tension brane denoted by$+$ ($-$
is the negative tension brane).
We have defined the field strength
\begin{equation}
F=dC
\end{equation}
and the dual field $*F$ is a scalar, hence the wedge product is not necessary in the kinetic terms.
The equations of motion are given by
\begin{equation}
\partial_5 (*F)= \frac{2\kappa_5}{\sqrt 3}Q(\delta_0-\delta_R)
\end{equation}
implying that
\begin{equation}
*F=\frac{\kappa_5}{\sqrt 3}Q\epsilon ({x_5}) .
\end{equation}
By duality we find that
\begin{equation}
F=\frac{\kappa_5}{\sqrt 3}Q\epsilon ({x_5})\sqrt{-g_5}dx^1\wedge
dx^2\wedge dx^3\wedge dx^4\wedge dx^5 .
\end{equation}
Notice that $F$ is odd on the orbifold implying that its integral, i.e. the boundary terms, in the action vanishes identically.
Substituting in the action and integrating over the extra--dimension we find
\begin{equation}
-\frac{\kappa_5^2 Q^2}{12 k }\int d^4 x \sqrt{-g} (1-e^{-4kR}) .
\end{equation}
In the Einstein frame, this leads to a potential for the radion
\begin{equation}
V_{F}=\frac{\kappa_5^2 Q^2}{12 k }\cosh 2\phi .
\end{equation}
The total potential energy including both the effects of detuning the brane tensions and the coupling to the bulk four--form reads
\begin{equation}
V=(\frac{\kappa_5^2 Q^2}{12 k }-\frac{3(1-T)k}{\kappa_5^2})\cosh 2\phi
\end{equation}
We find that there is a BPS bound
\begin{equation}
\vert Q_{BPS}\vert = \frac{6(1-T)k}{\kappa_5^2}
\end{equation}
where the charge of the brane is equal to the deficit of brane
tension. This case might be a necessary condition to allow the
supersymmetrization of the 4-form action, once proper
superpartners have been introduced, although we have not tried to
ascertain this result. Below the BPS bound, the radion potential
is unbounded from below. When the charge saturates the BPS bound,
we see immediately that the radion potential vanishes altogether,
i.e. the radion remains a flat direction. This signals the absence
of force between the branes. Above the BPS bound the radion
acquires a positive potential. The radion rolls down towards
infinity where the branes are infinitely far apart. Doing so, the
asymptotic dynamics are dominated by a cosmological constant. In
conclusion, the brane system is not stabilised once supersymmetry
is broken. In the most favourable case above the BPS bound, the
two branes repulse each other.

\section{Conclusion}

We have presented results on the spontaneous breaking of
supersymmetry in five dimensional supergravity with two
boundaries. We have studied   classical configurations
corresponding to rotated branes with  either singular intersecting
branes or parallel branes preserving $1/4$ of supersymmetry. We
have then analysed the low energy effective action involving the
coupling of the radion field to 4d gravity. The radion acquires a
negative and unbounded from below potential. The radion potential
can be rendered bounded from below by coupling the boundary branes
with a four--form. Above a BPS bound, the potential admits a
minimum corresponding to infinitely separated branes. When
saturating  the BPS bound linking the charge to the deficit of
tension,
 the radion has a flat potential.
Above the BPS bound the branes roll away to infinity where the remaining
cosmological constant is positive. The phenomenology of such a system is currently under study.

\section{Acknowledgments}

We would like to thank C. van de Bruck, C. Charmousis, C. Grojean, Z. Lalak,
J. Mourad and F. Quevedo for useful discussions and comments.

\section{Appendix A: Rotated branes and $AdS_4$ branes}

Consider a single detuned brane $T<1$. One can find a solution of the equations
of motion where slices of constant $z$ admit an $AdS_4$ metric
\begin{equation}
ds^2=a^2(z) g_{\mu\nu}d\tilde x^\mu d\tilde x^\nu + dz^2
\end{equation}
where $g_{\mu\nu}$ is the $AdS_4$ metric with cosmological constant
\begin{equation}
\Lambda_{AdS_4} = 3(T^2-1) k^2 <0
\end{equation}
and
\begin{equation}
g_{\mu\nu}d\tilde x^\mu d\tilde x^\nu=
e^{-2\sqrt{-\frac{1}{3}\Lambda_{AdS_4} }\tilde x_3}(-d\tilde
t^2+d\tilde x_1^2 +d\tilde x_2^2)+d\tilde x_3^2 .
\end{equation}
The scale factor reads
\begin{equation}
a(z)= \frac{\sqrt{-\frac{1}{3}\Lambda_{AdS_4}}}{k}\cosh (k z -C) .
\end{equation}
The metric is isometric to the warped metric of $AdS_5$
\begin{equation}
ds^2=e^{-2kx_5}\eta_{\mu\nu}d x^\mu d x^\nu + dx_5^2
\end{equation}
 using
\begin{eqnarray}
 x_3&=& \sqrt{-\frac{3}{\Lambda_{AdS_4}}}\tanh (C-k z)e^{\sqrt{-\frac{1}{3}\Lambda_{AdS_4}} \tilde x_3}\nonumber \\
x_5&=&\sqrt{-\frac{\Lambda_{AdS_4}}{3}}\frac{\tilde x_3}{k}-\frac{\ln(\frac{-\Lambda_{AdS_4}}{3k}\cosh(C-k z))}{k}\nonumber \\
\end{eqnarray}
and $\tilde t=t,\ \tilde x_1=x_1,\ \tilde x_2= x_2$.
The boundary condition at $z=0$ for the single brane reads
\begin{equation}
\tanh (C) = T .
\end{equation}
Under the previous mapping, the brane at $z=0$ is mapped to a rotated brane
\begin{equation}
y= h x_3
\end{equation}
in conformal coordinates $ky=e^{kx_5}$. Now let us try to include
a second brane with negative tension at $z=\rho$, the boundary
condition reads
\begin{equation}
\tanh (k\rho -C)= -T
\end{equation}
whose only solution is $\rho=0$. Hence the metric with $AdS_4$ slicing
at constant $z$ is not a solution of the two brane system.

\section{Appendix B: Massless Gravitons}
Let us give a thorough analysis of the Israel matching conditions
applied to the massless gravitons. We choose to work in the transverse traceless gauge where
\begin{equation}
n_{\mu\nu}(y,x^\mu)=\phi (y) e^{ip.x} n_{\mu\nu} .
\end{equation}
The condition (\ref{israel}) reduces to
\begin{equation}
h_{\lambda}(p_\mu n^\lambda_\nu +p_\nu n^\lambda_\mu -p^\lambda
n_{\mu\nu})= ak \frac{h_\mu n^{\mu\nu} h_{\nu}}{1+h^2}
(\eta_{\mu\nu}+h_{\mu}h_\nu) .
\end{equation}
Contracting with $p^\mu$ and using $p^2=0$ one finds that
\begin{equation}
h_\mu n^{\mu\nu} h_{\nu}=0
\label{q1}
\end{equation}
together with
\begin{equation}
h_{\lambda}(p_\mu n^\lambda_\nu +p_\nu n^\lambda_\mu -p^\lambda
n_{\mu\nu})=0 .
\label{bou}
\end{equation}
This equation can be treated separating two different cases.

\subsection{$p.h\ne 0$}

Let us choose two space--like orthogonal unit vectors $e^\mu_i,\
i=1,2$ , which are orthogonal to $p$ and $h$. This completes a
basis of 4d vectors. Projecting the boundary equation (\ref{bou})
leads to
\begin{equation}
e_i.n.e_j=0 .
\label{q2}
\end{equation}
Let us now choose a particular basis of traceless and symmetric tensors
comprising $p^\mu p^\nu$, $(h^\mu h^\nu -\frac{h^2}{4}\eta^{\mu\nu})$,
$(e_i^\mu e_i^\nu -\frac{1}{4}\eta^{\mu\nu})$, $ \frac{1}{2}(e_1^\mu e^\nu_2 +
e_1^\nu e^\mu _2)$, $ \frac{1}{2}( e^\mu _i h^\nu+ e_i ^\nu h^\mu)$
and $\frac{1}{2}( e^\mu _i p^\nu+ e_i ^\nu p^\mu)$.
Using the transverse condition and (\ref{q1},\ref{q2}), one finds that
\begin{equation}
n^{\mu\nu}=\sum_{i=1}^2   \frac{\alpha_i}{2}( e^\mu _i p^\nu+ e_i
^\nu p^\mu ) .
\end{equation}
Using the residual gauge invariance with vector $\epsilon^\mu e^{ip.x}$ such that
\begin{equation}
\epsilon^\mu= \sum_{i=1}^{2}\beta_i e_i^\mu
\end{equation}
 the gauge transformation
\begin{equation}
n_{\mu\nu}\to n_{\mu\nu}+ p_\mu\epsilon_\nu +p_\nu\epsilon_\mu
\end{equation}
one can fix $\beta_i=-\frac{1}{2} \alpha_i$.
Hence no massless graviton appears in the spectrum.

\subsection{$p.h=0$}
Notice first that one can choose a basis of vectors such that $e_i^\mu$
are now such that $h.e_i=0$, $e_1.e_2=0$ and $p.e_2=0$. Notice that $p.e_1\ne 0$ here. The new basis of traceless symmetric tensors is now $p^\mu p^\nu$, $(h^\mu h^\nu -\frac{h^2}{4}\eta^{\mu\nu})$,
$(e_i^\mu e_i^\nu -\frac{1}{4}\eta^{\mu\nu})$, $ \frac{1}{2}(e_1^\mu e^\nu_2 +
e_1^\nu e^\mu _2)$, $ \frac{1}{2}( e^\mu _i h^\nu+ e_i ^\nu h^\mu)$, $\frac{1}{2}( e^\mu _2 p^\nu+ e_2 ^\nu p^\mu)$ and $\frac{1}{2}(p^\mu h^\nu+p^\nu h^\mu)$.
The boundary equation reads now
\begin{equation}
h_{\lambda}(p_\mu n^\lambda_\nu +p_\nu n^\lambda_\mu)=0
\end{equation}
leading  to
\begin{equation}
e_i^\mu n_{\mu\nu}h^\nu=0 .
\end{equation}
The solutions to these constraints read  now
\begin{equation}
n_{\mu\nu}=  \alpha p_\mu p_\nu +\frac{\alpha_2}{2}( e^\mu _2
p^\nu+ e_2 ^\nu p^\mu ) .
\end{equation}
The residual gauge invariance parametrized by $\epsilon^\mu e^{ip.x}$ involves two  components
\begin{equation}
\epsilon^\mu= \beta_2 e_2^\mu + \beta p^\mu .
\end{equation}
Using $\beta=-\frac{1}{2}\alpha$ and $\beta_2=-\frac{1}{2}\alpha_2$,  no graviton
remains in the spectrum.

\end{document}